\documentclass[angeo]{copernicus}		

\graphicspath{{./}{catalog_25/}}

\usepackage{etoolbox}
\makeatletter
\patchcmd\@combinedblfloats{\box\@outputbox}{\unvbox\@outputbox}{}{\errmessage{\noexpand patch failed}}
\makeatother

\begin{document}

\title{Catalog of fine-structured electron velocity distribution functions -- Part 1: Antiparallel magnetic-field reconnection (Geospace Environmental Modeling case)}
\runningtitle{Catalog of electron velocity distributions -- Part 1: GEM case}
\runningauthor{Ph.-A. Bourdin}

\Author{Philippe-A.}{Bourdin}
\affil{Space Research Institute, Austrian Academy of Sciences, Schmiedlstr. 6, 8042 Graz, Austria}
\correspondence{Philippe-A. Bourdin (Philippe.Bourdin@oeaw.ac.at)}

\received{29 November 2016}
\revised{16 May 2017}
\accepted{31 July 2017}
\published{5 September 2017}

\firstpage{1051}
\pubvol{35}
\pubyear{2017}
\texlicencestatement{This work is distributed under\\the Creative Commons Attribution 3.0 License.}
\keywords{Magnetospheric physics (magnetotail) -- space plasma physics (magnetic reconnection; numerical simulation studies)}
\maketitle

\begin{abstract}
To understand the essential physics needed to reproduce magnetic reconnection events in 2.5-D particle-in-cell (PIC) simulations, we revisit the Geospace Environmental Modeling (GEM) setup.
We set up a 2-D Harris current sheet (that also specifies the initial conditions) to evolve the reconnection of antiparallel magnetic fields.
In contrast to the GEM setup, we use a much smaller initial perturbation to trigger the reconnection and evolve it more self-consistently.
From PIC simulation data with high-quality particle statistics we study a symmetric reconnection site, including separatrix layers, as well as the inflow and the outflow regions.
The velocity distribution functions (VDFs) of electrons have a fine structure and vary strongly depending on their location within the reconnection setup.
The goal is to start cataloging multidimensional fine-structured electron velocity distributions showing different reconnection processes in the Earth's magnetotail under various conditions.
This will enable a direct comparison with observations from, e.g., the NASA Magnetospheric MultiScale (MMS) mission, to identify reconnection-related events.
We find regions with strong non-gyrotropy also near the separatrix layer and provide a refined criterion to identify an electron diffusion region in the magnetotail.
The good statistical significance of this work for relatively small analysis areas reveals the gradual changes within the fine structure of electron VDFs depending on their sampling site.
\end{abstract}

\introduction

The notion of magnetic reconnection was originally introduced by \citet{Giovanelli:1946} to the space and astrophysical plasma physics community in order to explain violent energy releases, such as solar flares and coronal mass ejections at the Sun.
Nowadays, magnetic reconnection is known to occur also in the Earth magnetosphere, in particular at the dayside magnetopause, the cusp region, and in the magnetotail.
Magnetic reconnection is the most likely mechanism to drive auroral substorms \citep{Russell-McPherron:1973}.
Various theoretical models have been proposed to explain the mechanism operating in magnetic reconnection.
Major examples are the viscous-type reconnection by \citet{Sweet:1958} and \citet{Parker:1957}, the slow-shock acceleration model by \citet{Petschek:1964}, and the discontinuity-compound model by \citet{Sonnerup:1970}.
A number of numerical simulations as well as remote and in situ observations have also been performed to understand the spatial and temporal development of the reconnection process \citep{Paschmann:2013,Karimabadi+al:2013,Treumann-Baumjohann:2013,Treumann-Baumjohann:2015}.
Magnetic reconnection is also observed in laboratory plasmas, yet it is difficult to reach the non-collisional regime in reconnection experiments and transfer the obtained results to a space-plasma environment; see reviews by \citet{Zweibel-Yamada:2009} and \citet{Yamada+al:2010}.

Magnetic reconnection requires the breakdown of the frozen-in magnetic field from the magnetohydrodynamic point of view.
The motion of the magnetic field lines is described by the induction equation, and solving this equation requires detailed knowledge of the electric fields in the plasma, particularly on the kinetic scales where individual particle motions (gyration, drift, wave--particle resonance) will be effective.
Therefore, the reconnection region is divided into distinct scales: macroscopic magnetohydrodynamic scales and microscopic kinetic scales around the reconnection site where individual particle species need to be treated.
In an approximation based on the two-fluid model of plasma, the electric field on the kinetic scales is evaluated by the generalized Ohm's law, although it is a rather simplified picture, neglecting the wave--particle interactions such as the cyclotron or Landau resonances \citep{Pritchett:2001}.
We evaluate the induction equation with the help of a reduced proton-to-electron mass ratio (which is about 1836 in reality) and the generalized Ohm's law.

In particular, when the current sheet thickness (measured by the gradient scale or inhomogeneity of the magnetic field) becomes comparable or even smaller than the particle gyroradius, the velocity distributions are no longer Maxwellian nor gyrotropic \citep{Hoshino+al:2001}.
Recent kinetic simulations show that the velocity distributions are indeed unique with various realizations: two-sided together with a triangular distribution\citep{Ng+al:2012} and also including a swirl distribution \citep{Bessho+al:2014,Shuster+al:2015}.
These non-gyrotropic velocity distributions are obtained from a particle-in-cell (PIC) simulation in a two-dimensional reconnection setup.
Qualitatively, the non-gyrotropic distributions can be understood as effects of electron meandering motions\citep{Horiuchi+Sato:1994}, acceleration through electric fields, and deflection by the magnetic field -- similar to the situation observed for ions by \cite{Nagai+al:2015}.

Earlier and recent kinetic studies indicate that the electron stress must be the most relevant effect at the center of the reconnection region in a steady state \citep{Horiuchi+Sato:1994,Hesse+Winske:1998,Pritchett:2001,Hesse+al:2011}.
We aim to associate the electron velocity distribution functions with various regions of magnetic reconnection.
To this end, we run a numerical experiment to generate magnetic reconnection following Geospace Environmental Modeling (GEM) \citep{Birn+al:2001,Pritchett:2001} and systematically characterize the 3-D electron velocity distribution functions, in particular how the distribution functions are non-gyrotropic, and where they occur.

Here we present a comprehensive catalog of non-gyrotropic electron velocity distribution functions that are relevant to magnetic reconnection.
Such simulation results are to be compared to Magnetospheric MultiScale (MMS) spacecraft observations from the magnetotail, similar to what was done on the dayside by \citep{Burch+al:2016a}.
The catalog can be used to sum up electron velocity distributions along the trajectory of a spacecraft in order to characterize the magnetic-field configuration that was crossed.
In particular, we describe a characteristic feature of the electron diffusion region within antiparallel field reconnection.
Additional such catalogs for different configurations are needed for a better understanding of observed electron velocity distribution functions (VDFs).

\section{PIC simulation and analysis}
To compare the GEM setup with magnetotail observations, we basically require an antiparallel magnetic-field configuration, which is implemented by a Harris current sheet \citep{Harris:1962}.
We add an initial perturbation to the in-plane magnetic-field components $B_{\rm{x}}$ and $B_{\rm{z}}$ to trigger the reconnection in the center of the simulation domain.

We use the open-source code ``iPic3D'' \citep{Markidis+al:2010}, which implements the GEM setup together with our changes\footnote{See download link in the ``Data Availability'' section.} described in Sect.\,\ref{S:setup}.
With a large number of super-particles in our simulation we have access to good statistics about particle parameters even for small analysis regions.
We use a total of $13.1$ million super-particles for each species (electrons and ions) per $d_{\rm{i}}^2$, where $d_{\rm{i}}$ is the inertial length of background ions (see Sect.\,\ref{S:setup}).

The simulation results comprise the particle and bulk velocities, the magnetic and electric fields in three components, as well as the full pressure tensor of electrons and protons.

\subsection{Initial parameters and boundary conditions}\label{S:setup}
We use $25$ as a proton-to-electron mass ratio.
We perform another simulation with a mass ratio of $100$ and an unchanged background magnetic field ($B_0$).
The main differences we find are as follows: (1) electron velocities are scaled up proportionally to the square root of the mass ratio; (2) the electron diffusion region shrinks in its ${\rm{z}}$ extent with the same proportionality; and (3) the electron velocity shear layer is located closer to the separatrix.
As a result, we find similar shapes of the electron velocity distributions when we consider the slight spatial displacement of the electron velocity shear layer (see Sect.\,\ref{S:snapshot}) towards the almost unchanged location of the separatrix.
Using the same computational resources, a higher mass ratio implies stronger PIC noise, and the fine structure in the electron VDFs becomes less significant.
Still, one should continue this study and provide catalogs for comparison with more realistic mass ratios.

The computational domain covers $25.6 \times 12.8\,d_{\rm{i}}^2$ with $512 \times 256$ grid points, where $d_{\rm{i}}$ is the ion inertial length for the background number density $n_{\rm{0}} = 0.2$ and the initial background magnetic field $B_{\rm{0}} = 0.05477$.
The resulting grid spacing is $\Delta\rm{r} = 0.05\,d_{\rm{i}}$.
The initial condition follows $B_{\rm{x}}(z) = B_{\rm{0}}\,{\rm{tanh}}(z/D_{\rm{0}})$ for the magnetic field and $n(z) = n_{\rm{0}} (1 + 5\,{\rm{cosh}}^{-2}(z/D_{\rm{0}}))$ for the number density with $D_{\rm{0}} = 0.5447\,d_{\rm{i}}$ as the half-thickness of the current sheet.


The initial thermal velocity is $v_{\rm{th,e}} \approx 0.072\,c$ for electrons and $v_{\rm{th,i}} \approx 0.032\,c$ for ions.
Together with $B_0 = 0.05$ and the mass ratio of $25$, this implies a plasma beta of $\beta_{\rm{e}} = 1/6$ and $\beta_{\rm{i}} = 5/6$ for electrons and ions, respectively.
The Debye length $\lambda_D$ relates to the grid spacing as $\lambda_D \approx 0.29 \Delta \rm{r}$.
The time step remains fixed at $\Delta t = 0.5 / \Omega_{\rm{i}}$, where $\Omega_{\rm{i}} = {\rm{e\,B_0 / m}}$ is the ion gyrofrequency determined by the charge $e$, the ion mass $m$, and the initial background magnetic field amplitude $B_0$.
In the ${\rm{x}}$ direction we use a periodic boundary condition, as well as conducting walls at the ${\rm{z}}$ boundaries.
${\rm{y}}$ is degenerated in our case and is hence quasiperiodic or invariant.

We tested to what extent guide fields of $B_{\rm{y}} = 0.1$, $1$, and $10\% B_0$ influence the results obtained from the GEM case that has no guide field.
We find that guide fields of up to $1\% B_0$ do not significantly change the obtained electron VDFs presented in this work.
For a larger guide field of $10\% B_0$, we do see significant variations and hence propose to continue this cataloging study with guide fields of $1\% B_0$ and more, departing from the exactly antiparallel magnetic-field configuration in a stepwise way.

We use another simulation run with a quadruple box size of $51.2 \times 25.6\,d_{\rm{i}}^2$ to verify that we have no significant influence due to the domain boundaries in the results of this work.
Reflected particles from the ${\rm{z}}$ boundaries mainly propagate along the magnetic field, which is horizontal (mainly oriented along the ${\rm{x}}$ direction near the ${\rm{z}}$ boundaries) and does not yet connect to any of our regions of interest.
Particles that cross the ${\rm{x}}$ boundaries penetrate only the outer simulation domain and do not reach closer than about ${\rm{x}} = \pm 9\,d_{\rm{i}}$ to the reconnection center in significant quantities.
An example of such particles traveling inwards is visible as a minor third peak in $v_{\rm{x}} > 0.1\,c$ at the position of ${\rm{x}} = 9$ and ${\rm{z}} = 0.6\,d_{\rm{i}}$; see the video in the Supplement.\footnote{\url{https://doi.org/10.5446/31796}}

Because we need to provide the statistical significance in order to obtain low-noise fields and fine-structured electron velocity distributions, we require an unprecedentedly large number of particles (in total $8.6$ billion).
When scaling the mass ratio to higher values, typical spatial scales become smaller for the electrons, which requires having smaller analysis areas.
Hence, this either requires a currently unfeasible increase in computational demands while trying to maintain the data quality from this work for much higher mass ratios, or one has to change the original GEM parameters (like $B_0$), which would on the other hand prevents us from comparing this catalog directly with earlier GEM simulation results.
Also, the simulation run with a larger box size needs to be conducted with less particles per $d_{\rm{i}}^2$.
Therefore, for now we retain the original GEM settings, including the mass ratio and simulation box size to obtain the lowest possible PIC noise level.

\subsection{Comparison of reconnection rates}\label{S:reconnection_rate}
While the original GEM setup proposes a large perturbation (covering the whole simulation domain), we trigger the reconnection with a perturbation that is 10 times smaller in its spatial extent.
The effect of this modification is a smaller gas-to-magnetic pressure disequilibrium in the initial condition and a later onset of the reconnection.
The reconnected field topology becomes more symmetric because both peaks in the perturbation are closer to the middle of the simulation domain.
Therefore, we evolve the reconnection in a more self-consistent way.

\begin{figure}[t]
\includegraphics[width=8.25cm]{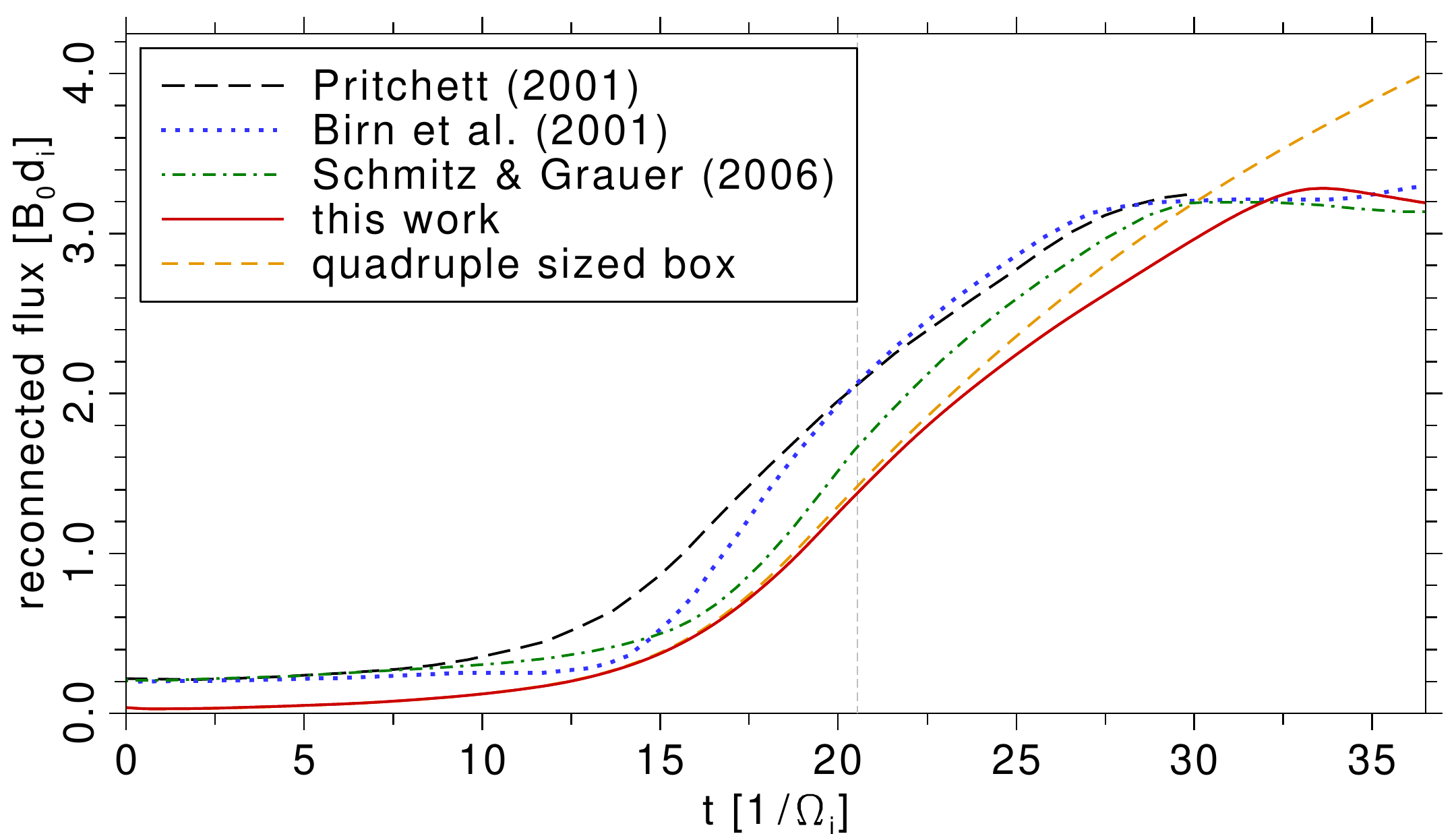}
\caption{Evolution of the reconnected flux in our reference model, as well as for a quadruple-sized simulation domain, together with the data from \cite{Pritchett:2001}, \cite{Birn+al:2001}, and \cite{Schmitz+Grauer:2006}.
The vertical gray dashed line indicates the time of the data snapshot we analyze in this work.}
\label{F:reconnected_flux}
\end{figure}

The reconnection rate is the slope of the reconnected flux.
We find that our setup generates a similar reconnection rate as \cite{Pritchett:2001}, while the reconnected flux is different due to the initial condition; see Fig.\,\ref{F:reconnected_flux}.
If we subtract the difference of both initial conditions and consider that the onset of the reconnection is about $\Delta t = 2\,\Omega_{\rm{i}}^{-1}$ later in our simulation, both curves become very similar.
A later onset was also observed by \cite{Birn+al:2001}, while the reconnection rate evolves differently in their work.
\cite{Schmitz+Grauer:2006} used a Vlasov code to model the GEM setup, and they also find a later onset.
Their reconnection rate lies in between the ones observed by \cite{Birn+al:2001} and \cite{Pritchett:2001}.

To check if the simulation domain size has a significant impact on the reconnected flux, we compare it with a simulation run that has a quadruple-sized box.
All other parameters and the initial conditions are kept identical.
We find an almost identical evolution of the reconnected flux until $t = 22\,\Omega_{\rm{i}}^{-1}$.
After that time, the reconnected fluxes start to deviate from each other (see the dashed red and orange lines in Fig.\,\ref{F:reconnected_flux}).
We find that the plateau seen in \cite{Pritchett:2001}, \cite{Birn+al:2001}, and \cite{Schmitz+Grauer:2006} after $t=27\,\Omega_{\rm{i}}^{-1}$ is caused by the box size and is hence due to influence from the boundary conditions.
Until $t = 20.54\,\Omega_{\rm{i}}^{-1}$, we see no significant influence from the box boundaries.
Therefore, we use this snapshot for our analysis and refrain from using data at later times.

From our lager-box simulation run, we see that the magnetic influx and hence the reconnection processes are slowly being suppressed and come to a halt after $t = 33\,\Omega_{\rm{i}}^{-1}$; see the differences between the red solid and the orange dashed line in Fig.\,\ref{F:reconnected_flux}.

We note that the comparability between different simulations is limited because we basically watch at different evolution stages of the reconnection when we check for identical times; see the vertical gray dashed line in Fig.\,\ref{F:reconnected_flux}.
As compared to our results, the amount of total reconnected flux at this time is about 52\,\% higher in \cite{Birn+al:2001,Pritchett:2001} and 21\,\% higher in \cite{Schmitz+Grauer:2006}.
Therefore, not the time of the data snapshot but instead the amount of reconnected flux (without the initial perturbation) is the better quantity to compare between simulation works.

\subsection{Evolved reconnection}\label{S:snapshot}
\begin{figure}[t]
\includegraphics[width=8.05cm]{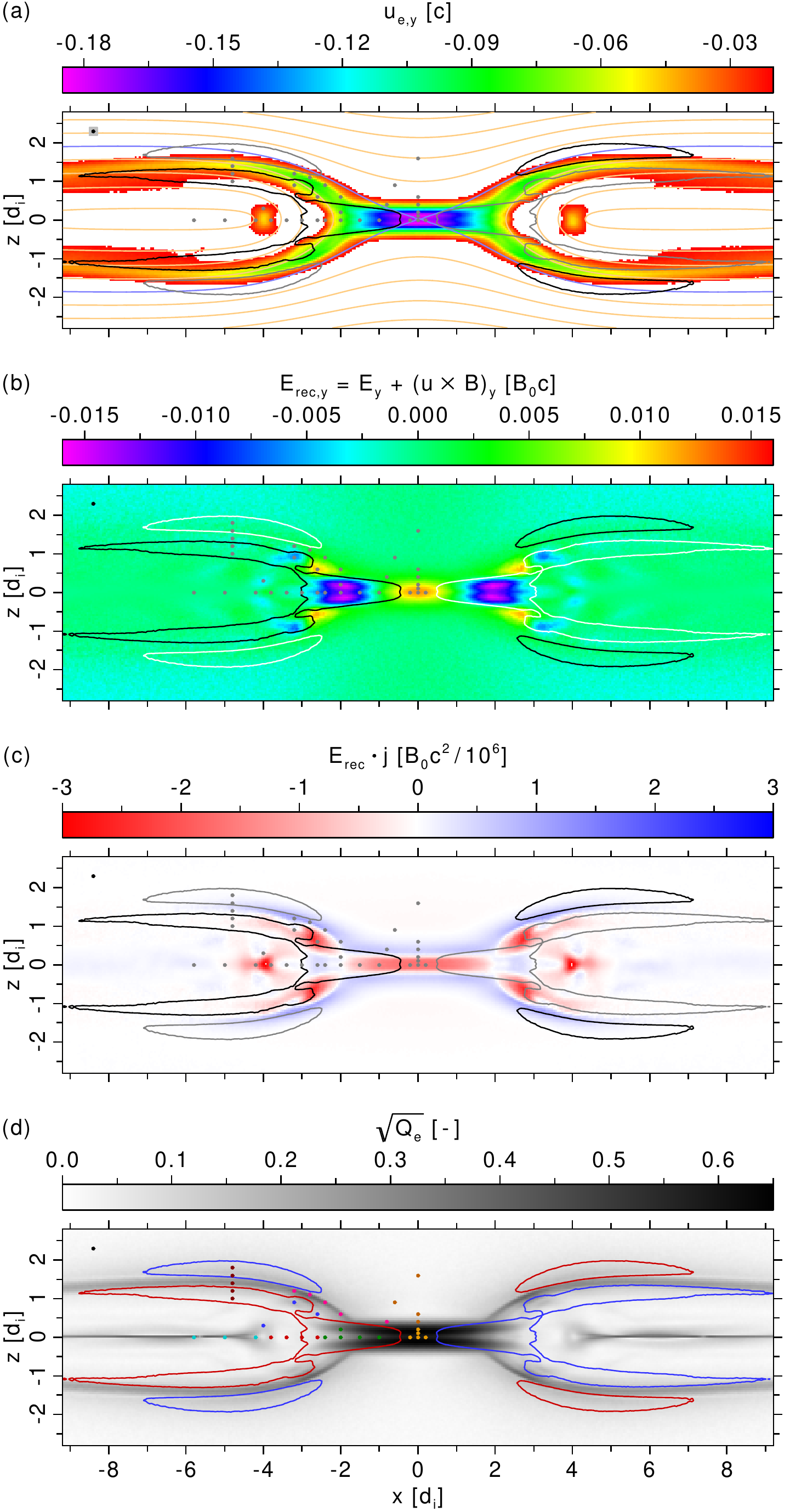}
\caption{Magnetic reconnection snapshot with indications of the sampling points for the electron velocity distribution functions.
The coordinate system is right-handed; Earth is to the right, and $\rm{y} > 0$ points towards dusk.
Panel~\textbf{(a)} visualizes the electron bulk velocity in the out-of-plane component $u_{\rm{e,y}}$ (color-coded).
The gray square indicates the size of the analysis area around the analysis positions (dots).
Overplotted orange lines follow the in-plane magnetic field, while the field line crossing the reconnection center is blue.
The contour lines indicate a speed of $u_{\rm{e,x}} = \pm 0.04\,{\rm{c}}$ in all panels, where black or red represents a leftwards flow and white or blue a rightwards flow.
In panel~\textbf{(b)} we show the out-of-plane component of the reconnection electric field $\vec{E_{\rm{rec}}} = \vec{E} + \vec{u} \times \vec{B}$ (color-coded).
Panel~\textbf{(c)} contains the scalar product of the reconnection electric field and the current density $\vec{E_{\rm{rec}}} \cdot \vec{j}$ (coded in red, white, and blue).
Panel~\textbf{(d)} displays the square root of the non-gyrotropy index $\sqrt{Q_{\rm{e}}}$ (linear grayscale).
The color code of the dots indicates the grouping of analysis regions.}
\label{F:snapshot_overview}
\end{figure}

We display a snapshot of the reconnection in Fig.\,\ref{F:snapshot_overview}, where the amount of flux that has reconnected after the initial condition is $1.39\,B_0$.
The simulation time here is $t = 20.54\,\Omega_{\rm{i}}^{-1}$.
We obtain the reconnection center exactly in the middle of the simulation box, where the out-of-plane component of the reconnection electric field $E_{\rm{rec}} = \vec{E} + \vec{u} \times \vec{B}$ is dominant.

To estimate the non-gyrotropy of electrons we use the index proposed by \cite{Swisdak:2016} (Eqs.\,A5--A8) that we compute as
\[Q_{\rm{e}} = 1 - 4 I_2 / [(I_1 - P_\parallel) (I_1 + 3 P_\parallel)]\]
where $I_1$ is the trace of the pressure tensor $\mathbb{P}$, its field-parallel component is $P_\parallel = \vec{e_B}^T \mathbb{P} \vec{e_B}$ with a unit vector along the magnetic field $\vec{e_B}$, and $I_2$ is
\[I_2 = P_{\rm{xx}} P_{\rm{yy}} + P_{\rm{xx}} P_{\rm{zz}} + P_{\rm{yy}} P_{\rm{zz}} - (P_{\rm{xy}} P_{\rm{yx}} + P_{\rm{xz}} P_{\rm{zx}} + P_{\rm{yz}} P_{\rm{zy}})\]

The region with a high non-gyrotropy index $Q_{\rm{e}}$ roughly follows the reconnection current sheet, which is strongest at ${\rm{x}} = 0$ and ${\rm{z}} = \pm 0.15\,d_{\rm{i}}$ and is elongated along the mean background magnetic field direction ${\rm{x}}$.

Also, we find an increased value of $Q_{\rm{e}}$ that follows the electron velocity shear layer close to the separatrix; see green and yellow encoded $u_{\rm{e,y}}$ located around ${\rm{z}}= \pm 1.4\,d_{\rm{i}}$ in Fig.\,\ref{F:snapshot_overview}a.
This shear layer is enclosed by an electron bulk that is accelerated away from (towards) the reconnection site along the ${\rm{x}}$ direction; see the black/red (white/blue) contour lines in Fig.\,\ref{F:snapshot_overview} that correspond to downstreaming (upstreaming) electrons, respectively.
Exactly in between these ${\rm{x}}$ shear flows we see a strongly enhanced out-of-plane bulk velocity along the positive ${\rm{y}}$ direction.
There, the non-gyrotropy index $Q_{\rm{e}}$ is strongly enhanced even outside the electron diffusion region and in the absence of a significant out-of-plane electric field; see Fig.\,\ref{F:snapshot_overview}b and c.

\begin{figure}[t]
\includegraphics[width=8.25cm]{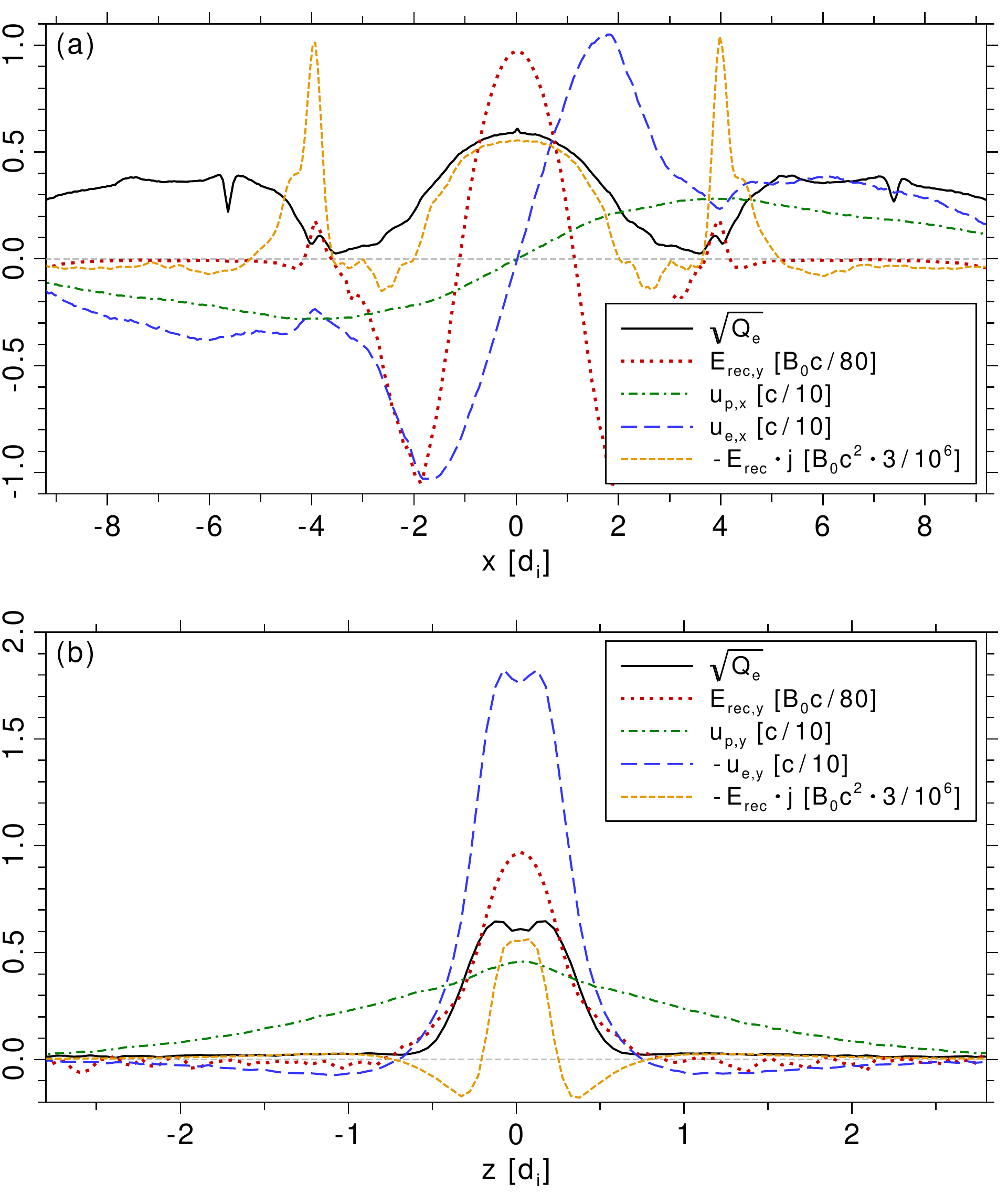}
\caption{Cuts for the non-gyrotropy index $Q_{\rm{e}}$, the reconnection electric field $E_{\rm{rec}}$, the scalar product the of total electric field and current density $\vec{E_{\rm{rec}}} \cdot \vec{j}$, and the bulk velocities $u$, along the ${\rm{z}}=0$ line~\textbf{(a)} and along ${\rm{x}}=0$~\textbf{(b)}.}
\label{F:cuts}
\end{figure}

In Fig.\,\ref{F:cuts} we show cuts of $Q_{\rm{e}}$ and $E_{\rm{rec}}$ together with the electron and ion bulk velocities $u_{\rm{e}}$ and $u_{\rm{i}}$ along the ${\rm{x}}$ and ${\rm{z}}$ axes.
In the cut along ${\rm{x}}$ we find a peak in the non-gyrotropy index $Q_{\rm{e}}$ in the reconnection center, where $E_{\rm{rec}}$ is strongest.
However, also in regions away from the center $\sqrt{Q_{\rm{e}}}$ is significantly enhanced and reaches values above 0.3 where $E_{\rm{rec}}$ is practically not present.
Therefore, we need another indicator besides the non-gyrotropy index in order to identify an electron diffusion region.

Along the $z$ direction, the bulk velocity $u_{\rm{e}}$ and $\sqrt{Q_{\rm{e}}}$ both have a double-peak shape, while the reconnection electric field $E_{\rm{rec}}$ clearly has a single peak.
The separation distances of the double peaks in $\sqrt{Q_{\rm{e}}}$ and $u_{\rm{e,y}}$ differ by $0.1\,d_{\rm{i}}$.

We compare $\sqrt{Q_{\rm{e}}}$ \citep{Swisdak:2016} with the scalar product of the reconnection electric field and the current density $\vec{E_{\rm{rec}}} \cdot \vec{j}$ \citep{Zenitani+al:2011}.
This latter quantity was also used to identify the electron diffusion region in \cite{Burch+al:2016b}.
We find that these two quantities anticorrelate well along the ${\rm{x}}$ axis ($\rm{z} = 0$) within ${|\rm{x}|} < 2 d_{\rm{i}}$ near the electron diffusion region; see the orange dashed and the black lines in Fig.\,\ref{F:cuts}a.
This means that $Q_{\rm{e}}$ is large along $\rm{z} = 0$ and near the reconnection center, where the current density $\vec{j}$ is antiparallel to $\vec{E}$ so that their scalar product becomes negative.
This correlation breaks down for ${|\rm{x}|} > 5 d_{\rm{i}}$, where $\vec{E_{\rm{rec}}} \cdot \vec{j}$ vanishes but $\sqrt{Q_{\rm{e}}}$ is significantly high.
Also, the strong peaks in $\vec{E_{\rm{rec}}} \cdot \vec{j}$ at ${\rm{x}} = \pm 4 d_{\rm{i}}$ are not seen in $\sqrt{Q_{\rm{e}}}$.
On the other hand, along the ${\rm{z}}$ axis we simply do not see a similar anticorrelation as along the ${\rm{x}}$ axis; see Fig.\,\ref{F:cuts}b.

For completeness, we checked that the non-gyrotropy index $A${\O}$_{\rm{e}}$ \citep[as defined in][]{Scudder-Daughton:2008} gives similar results to $Q_{\rm{e}}$, except that $A${\O}$_{\rm{e}}$ shows a stronger relative enhancement near the separatrices as compared to the current sheet in the central reconnection region.
The electron diffusion region is also indicated as being slightly larger in $A${\O}$_{\rm{e}}$ along the ${\rm{z}}$ direction than compared to $Q_{\rm{e}}$.

Altogether, we do not find a clear correlation of $Q_{\rm{e}}$ with any other quantity plotted in Fig.\,\ref{F:cuts}.
This suggests that the off-diagonal elements of the pressure tensor increase also through processes that are subsequent to the electron acceleration from the reconnection electric field $E_{\rm{rec}}$ -- or that are, in other words, not directly induced in the central electron diffusion region.

\subsection{Electron velocity distribution functions}\label{S:distributions}
\begin{figure*}[t]
\includegraphics[width=10.15cm]{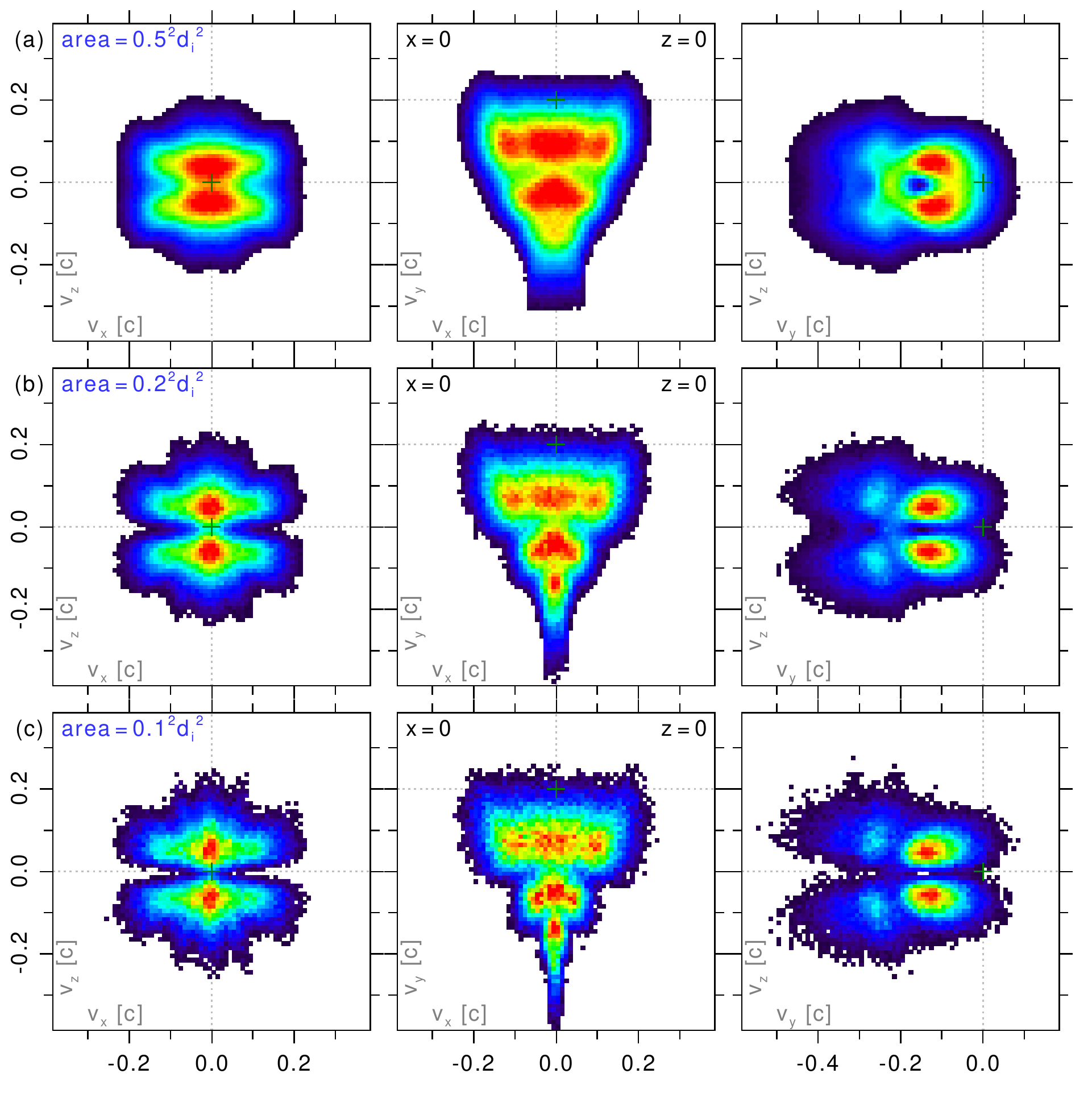}
\caption{Comparison of different analysis area sizes. Panel row~\textbf{(a)} is averaged similar to \cite{Zenitani-Nagai:2016}, \textbf{(b)}~reflects the quality presented in this work, and~\textbf{(c)} features an even smaller analysis area size; see Sect\,\ref{S:distributions}.}
\label{F:area-sizes}
\end{figure*}

We select small regions of interest with a size of $0.2 \times 0.2\,d_{\rm{i}}^2$, where we bin the electron velocities of particles contained in the region.
This size also reflects the electron gyro-radius where the magnetic field reaches $B_0/4$, like near the reconnection center.
\cite{Zenitani-Nagai:2016} average the electron velocity distributions over regions of $0.5 \times 0.5\,d_{\rm{i}}^2$, which results in less fine-structured electron velocity distributions; see Fig.\,\ref{F:area-sizes}.
Hence, we need to average over areas of $0.2 \times 0.2\,d_{\rm{i}}^2$ in order to capture distinct electron populations -- or to stay within about one electron gyration radius in a weak-field regime, like near the reconnection site.
Even smaller analysis regions of $0.1 \times 0.1\,d_{\rm{i}}^2$ would not reveal significant additional fine structures above the noise level; see row (c) in Fig.\,\ref{F:area-sizes}.

The differences between the row (a) in our Fig.\,\ref{F:area-sizes} and the original Fig.\,4, panels (e1)-(e3) in \cite{Zenitani-Nagai:2016} are due to the different time in the evolution of the reconnection process.
We see that the fine structure formed during the free evolution of the reconnection ($t = 20.54\,\Omega_{\rm{i}}^{-1}$), and it slowly decays or washes out when reaching the plateau phase (due to numerical constraints; see Fig.\,\ref{F:reconnected_flux}) at $t = 35\,\Omega_{\rm{i}}^{-1}$ that \cite{Zenitani-Nagai:2016} analyze.

We also average the non-gyrotropy index $Q_{\rm{e}}$ proposed by \cite{Swisdak:2016} within the analysis regions.
Each region contains about half a million super-particles per species.

\begin{figure*}[t]
\includegraphics[width=12.80cm]{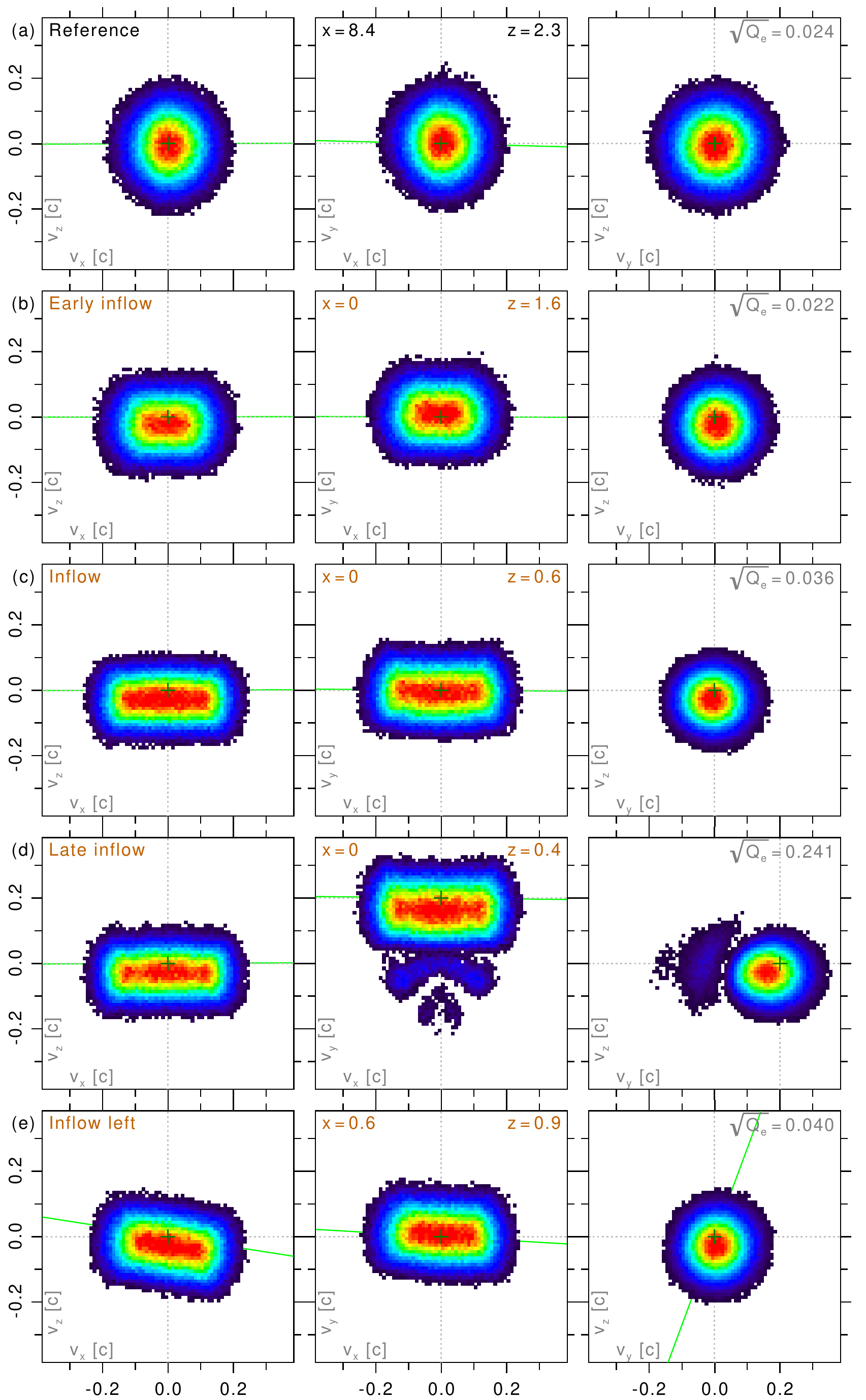}
\caption{Electron velocities distribution functions (2-D cuts) for the reference area and the inflow region.
The gray dotted line indicates $v = 0$ and the text label color indicates the analysis area positions from Fig.\,\ref{F:snapshot_overview}.
The ${\rm{z}}$ coordinates are given in $d_{\rm{i}}$ while ${\rm{x}}$ coordinates are given in negative $d_{\rm{i}}$; see the top of the panels in the middle column.
The color code shows the probability distribution on a linear scale, where saturated red indicates the $90\,\%$ level of the maximum value within each panel.
The green lines indicate the mean magnetic-field direction in those cases in which there is a significant one.}
\label{F:vel-dist_reference-inflow}
\end{figure*}

\begin{figure*}[t]
\includegraphics[width=12.97cm]{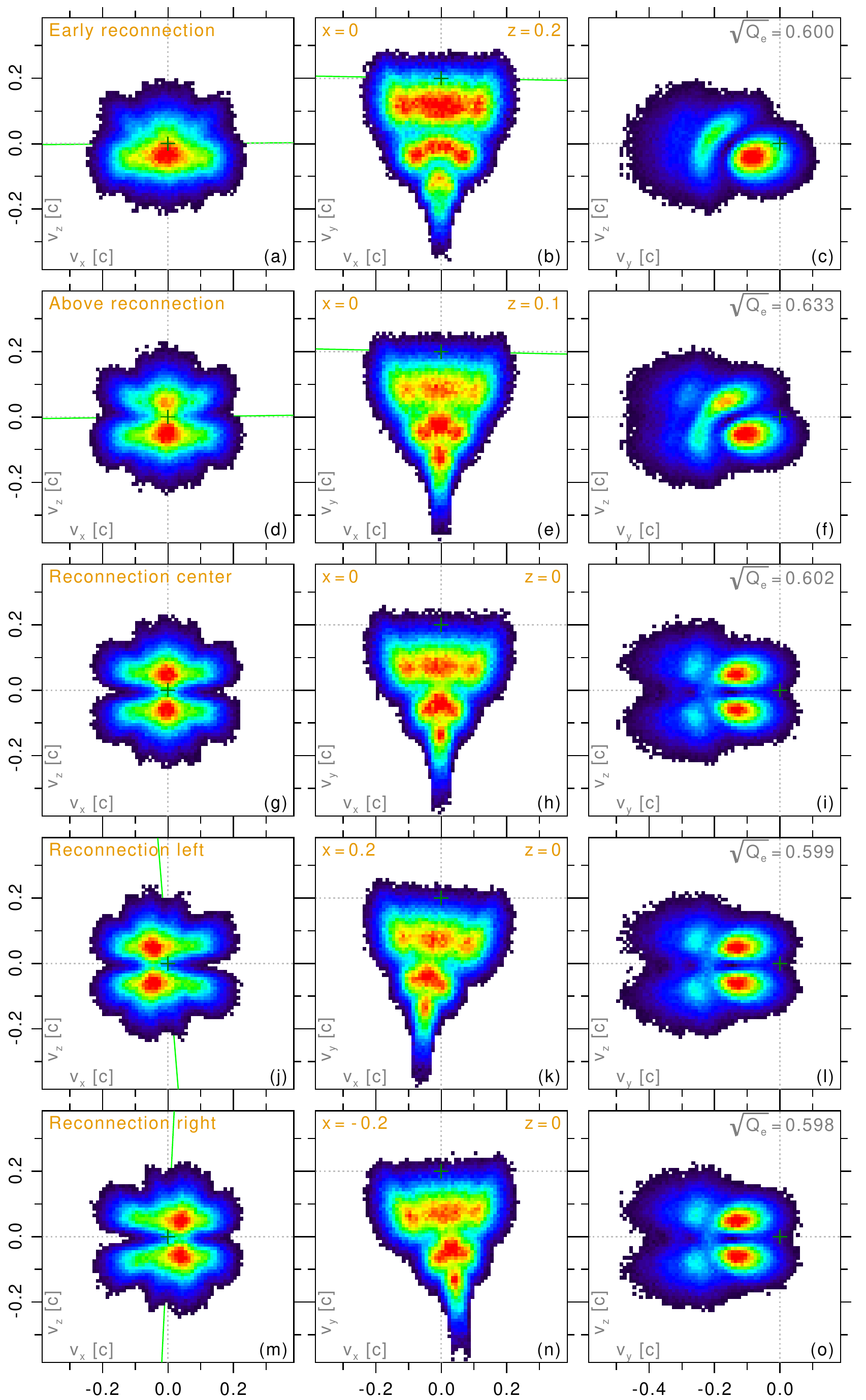}
\caption{Same as Fig.\,\ref{F:vel-dist_reference-inflow} but around the electron diffusion region.}
\label{F:vel-dist_reconnection}
\end{figure*}

\begin{figure*}[t]
\includegraphics[width=12.97cm]{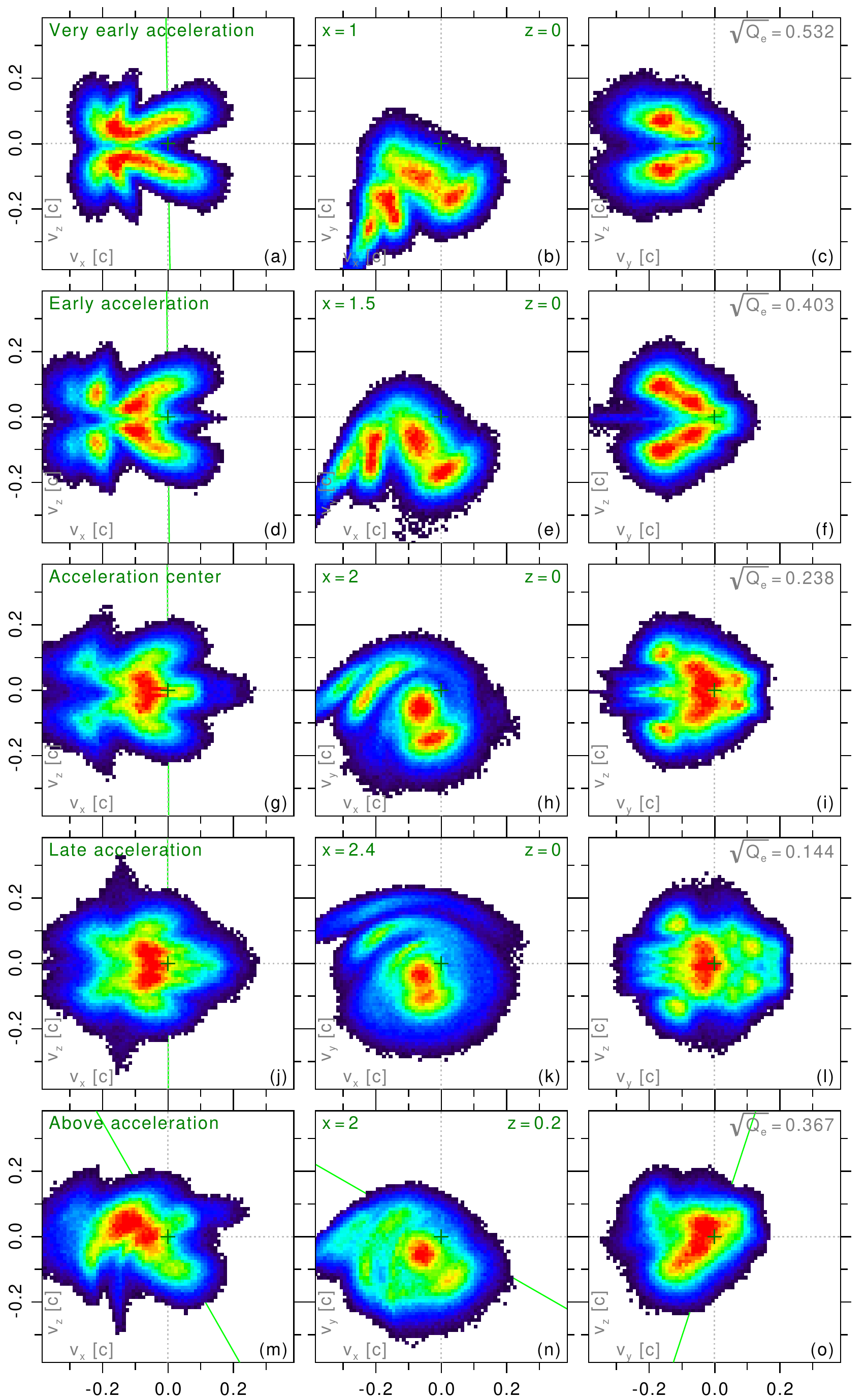}
\caption{Same as Fig.\,\ref{F:vel-dist_reference-inflow} but for the acceleration region.}
\label{F:vel-dist_acceleration}
\end{figure*}

\begin{figure*}[t]
\includegraphics[width=12.97cm]{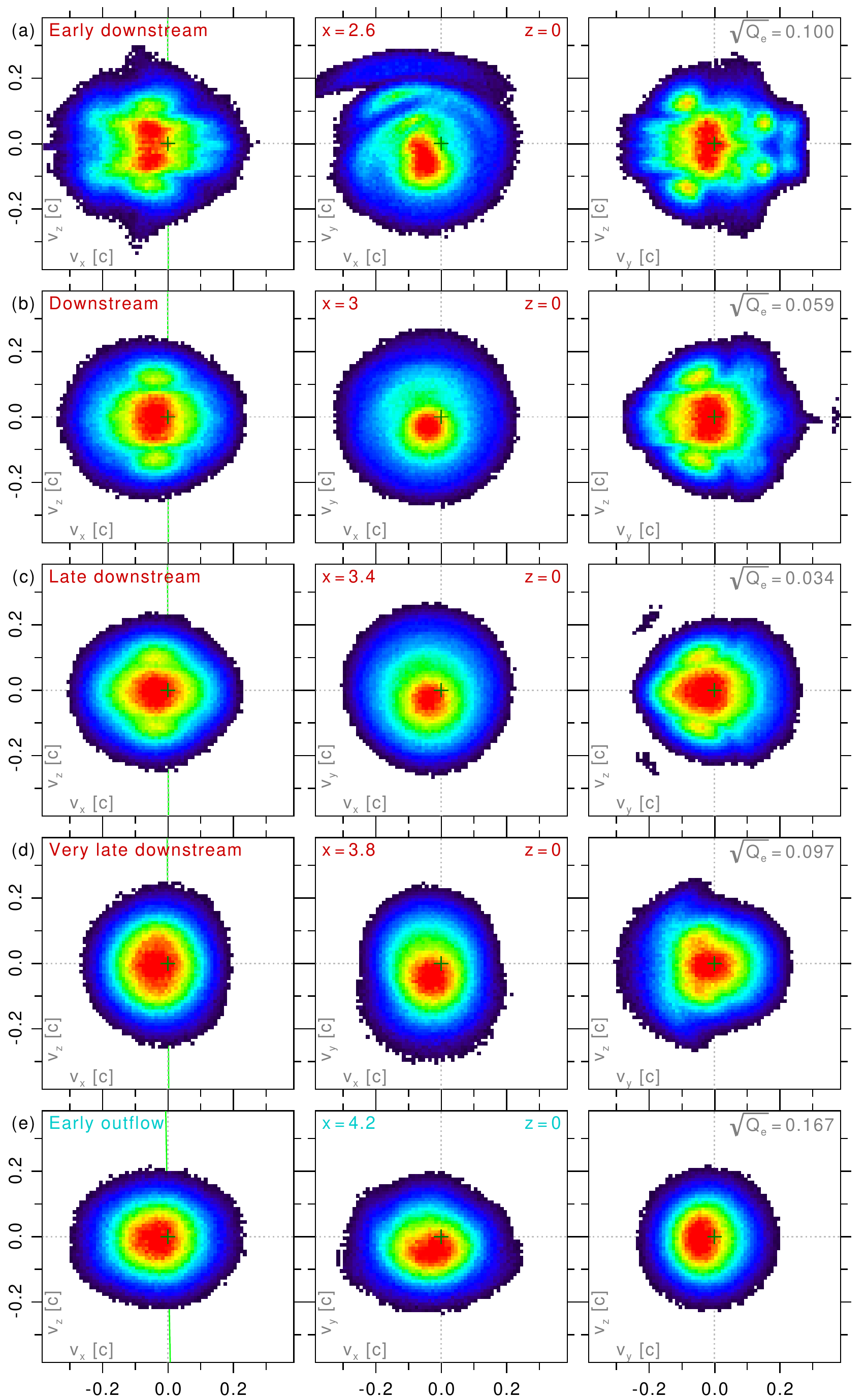}
\caption{Same as Fig.\,\ref{F:vel-dist_reference-inflow} but downstream of the acceleration and towards the outflow.}
\label{F:vel-dist_downstream}
\end{figure*}

\begin{figure*}[t]
\includegraphics[width=12.97cm]{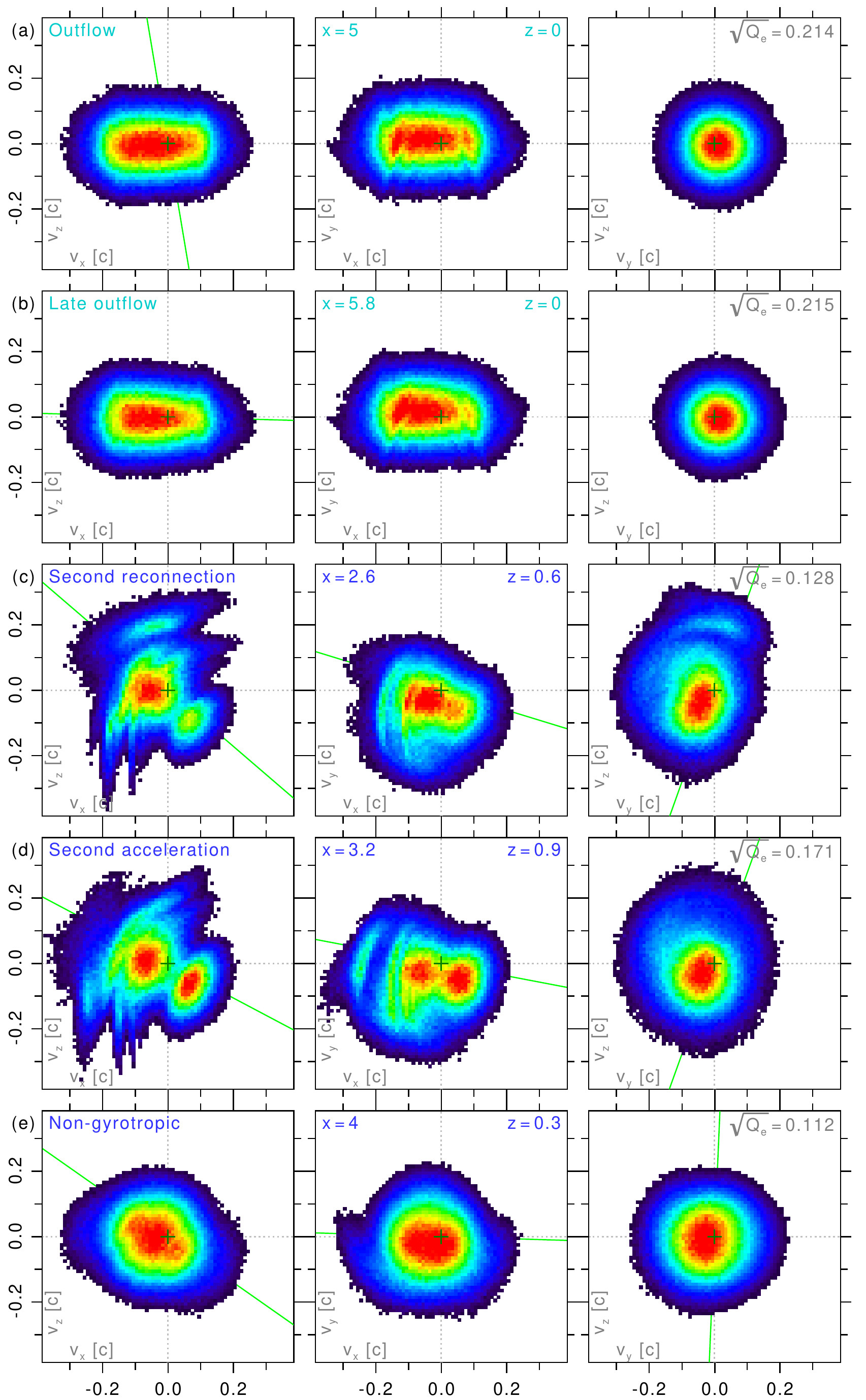}
\caption{Same as Fig.\,\ref{F:vel-dist_reference-inflow} but showing the outflow and regions with an enhanced non-gyrotropy index $Q_{\rm{e}}$.}
\label{F:vel-dist_outflow_high-Q}
\end{figure*}

\begin{figure*}[t]
\includegraphics[width=12.97cm]{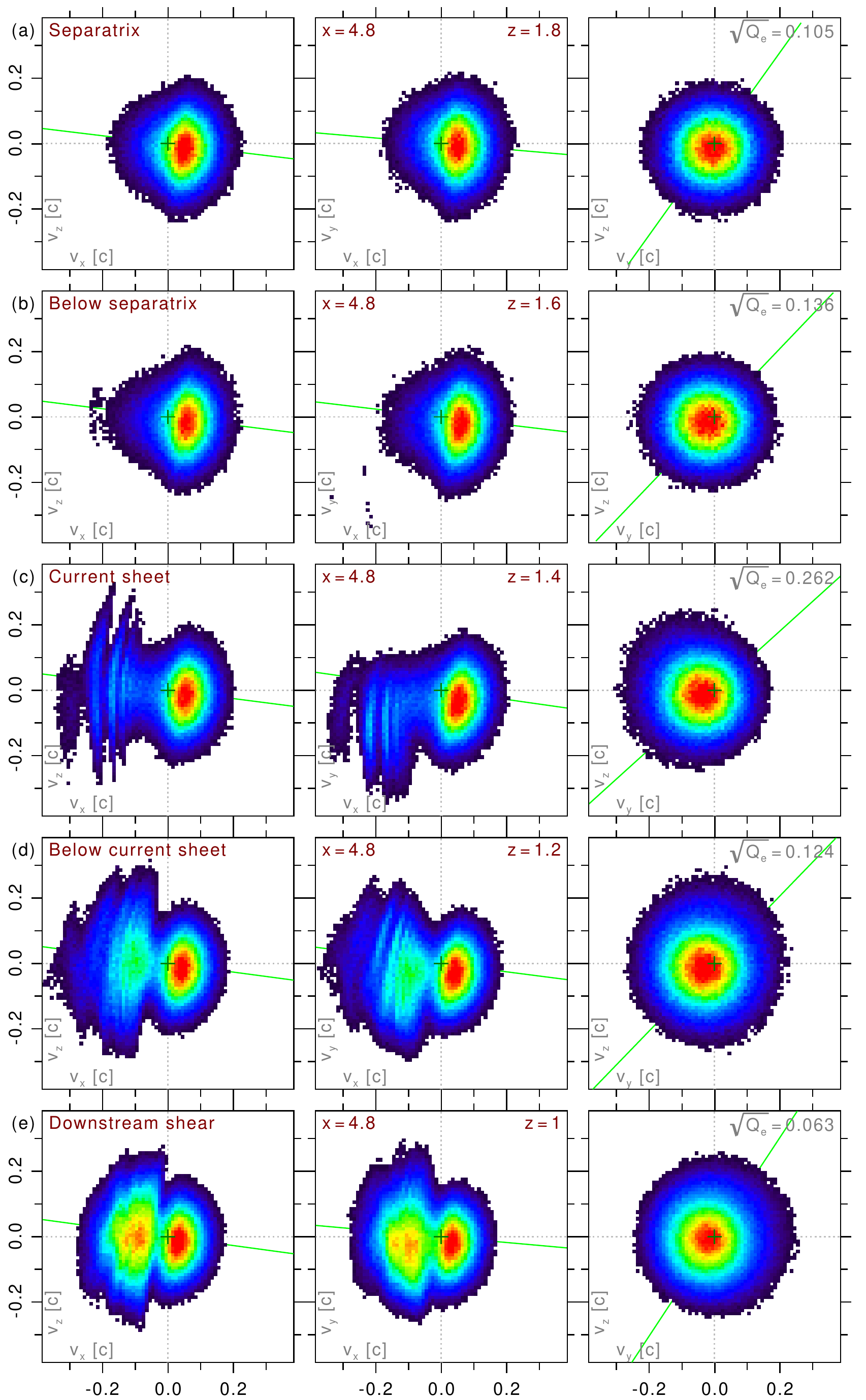}
\caption{Same as Fig.\,\ref{F:vel-dist_reference-inflow} but near the separatrix layer across an electron up-/downstream shear flow.}
\label{F:vel-dist_separatrix-shear}
\end{figure*}

\begin{figure*}[t]
\includegraphics[width=12.97cm]{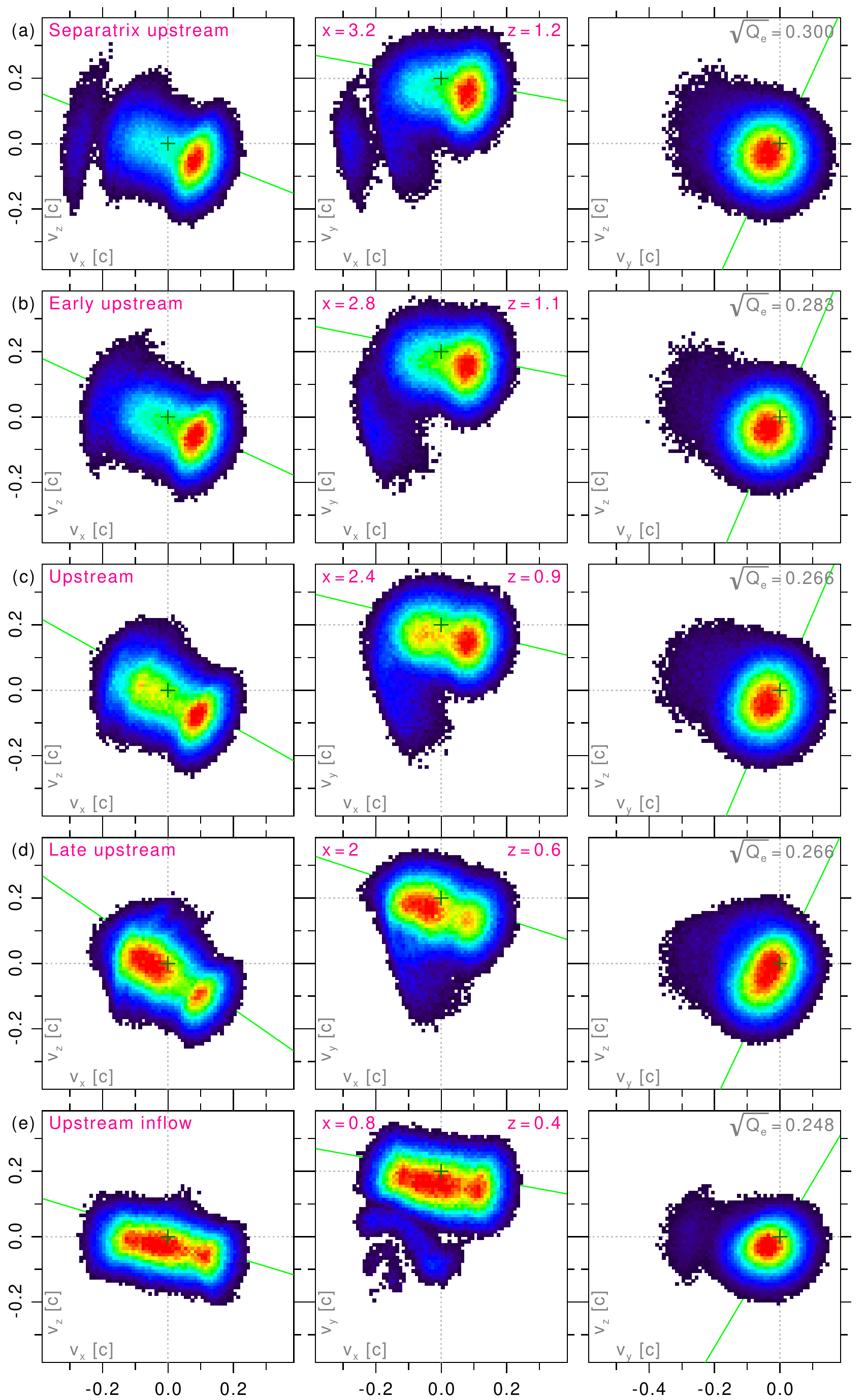}
\caption{Same as Fig.\,\ref{F:vel-dist_reference-inflow} but following the separatrix layer in the upstream direction towards the inflow.}
\label{F:vel-dist_upstream}
\end{figure*}

In Figs.\,\ref{F:vel-dist_reference-inflow} to \ref{F:vel-dist_upstream} we provide a comprehensive catalog of 3-D electron velocity distribution functions as 2-D cuts along the simulation coordinate directions and with integrated distributions along the direction orthogonal to each 2-D cut.
The text label color indicates the location of the analyzed regions group, cf. the colored dots in Fig.\,\ref{F:snapshot_overview}d.
The 2-D velocity distributions are along the mean initial magnetic field (${\rm{x}}$), the out-of-plane direction (${\rm{y}}$), and along the initial magnetic-field gradient (${\rm{z}}$, perpendicular to the central current sheet).

Fig.\,\ref{F:vel-dist_reference-inflow} contains the reference area (top row) that shows a clearly Maxwellian shape, represented by Gaussian distributions in all three directions.
The other panels in Fig.\,\ref{F:vel-dist_reference-inflow} are samples within the inflow region, where we find rectangular shapes in the \mbox{$v_{\rm{x}}$ -- $v_{\rm{z}}$} and the \mbox{$v_{\rm{x}}$ -- $v_{\rm{y}}$} plane, while the \mbox{$v_{\rm{y}}$ -- $v_{\rm{z}}$} remains Maxwellian, as also reported by \cite{Schmitz+Grauer:2006}.
The region labeled ``late inflow'' is already close to the reconnection center and features a gradually appearing triangular shape in the \mbox{$v_{\rm{x}}$ -- $v_{\rm{y}}$} cut.

When leaving the inflow region and approaching the reconnection site, we sample a single-peak shape along $v_{\rm{z}}$ in Fig.\,\ref{F:vel-dist_reconnection}.
At the reconnection center, we find that the inflowing electrons are accelerated by the reconnection electric field $E_{\rm{rec}}$ along the $\rm{y}$ direction, which forms the elongated tip of the triangular-shaped distribution in the \mbox{$v_{\rm{x}}$ -- $v_{\rm{y}}$} cut; see ``reconnection center'' row in Fig.\,\ref{F:vel-dist_reconnection}.
Also, we can confirm the increasing tilt of that tip when we consecutively sample regions on the ${\rm{x}}$ axis but with increasing distance from the reconnection center, as shown in \cite{Shuster+al:2015}.
This increasing tilt is due to the electric field vector that points perpendicular to the ${\rm{x}}-{\rm{y}}$ plane only in the exact reconnection center, but has a growing in-plane component when going away from the reconnection center.

The fine structure in the \mbox{$v_{\rm{x}}$ -- $v_{\rm{y}}$} panels (Fig.\,\ref{F:vel-dist_reconnection}b, e, h, k, and n) is caused by the number of electron meandering motions through oppositely oriented magnetic fields above and below the reconnection center.
This is very similar to the behavior of electrons for an antiparallel field plus a guide field, as described by \cite{Ng+al:2012}.
We also find gradual changes from $\rm{x}=\rm{z}=0$ (Fig.\,\ref{F:vel-dist_reconnection}, middle row) to $\rm{x}=1, 1.5, 2$; see the three upper rows in Fig.\,\ref{F:vel-dist_acceleration} (visible even better in the online movie).
It becomes clear that individual populations, like the red tip of the downwards-pointing triangle and the red v-shaped population above this tip in Fig.\,\ref{F:vel-dist_reconnection}h (two lowermost peaks), are the same populations as the green and orange distributions in the \mbox{$v_{\rm{x}}$ -- $v_{\rm{y}}$} panel shown in Fig.\,\ref{F:vel-dist_acceleration}h (two leftmost peaks) at $\rm{x}=2$, $\rm{z}=0$.

The strong red peaks in the center of the three lower rows of Fig.\,\ref{F:vel-dist_acceleration} are therefore from electrons that have not completed any full meandering motion, but that have basically evolved from the inflowing velocity distribution population directly; see row~(c) in Fig.\,\ref{F:vel-dist_reference-inflow}.

After crossing the reconnection center, we find a double-peak shape along $v_{\rm{z}}$ that is maintained until the plasma reaches the ``acceleration center'' (Fig.\,\ref{F:vel-dist_acceleration}) which has a similar spiral shape in the \mbox{$v_{\rm{x}}$ -- $v_{\rm{y}}$} plane, as also found by \cite{Bessho+al:2014}.

Figure~\ref{F:vel-dist_downstream} contains samples of the plasma exhaust downstream of the reconnection and acceleration regions.
The velocity distributions gradually become more gyrotropic again, as seen from the ``late downstream'' to the ``early outflow'' regions.
In particular, the differences in the work of \cite{Shuster+al:2015} between their panels rows F ($t=20$) and G ($t=29\,\Omega_{\rm{i}}^{-1}$) in their Fig.\,3 are explicable by the suppression of the reconnection process due to the box size (or the boundary conditions) at later times (cf. our Sect.\,\ref{S:reconnection_rate}).
While at $t=20$ their and our distributions are of course similar, we see that the previously created non-gyrotropic distributions downstream of the reconnection site decay with time.
In particular, these non-gyrotropic distributions are again approaching a more Maxwellian-like shape after $t=20$, which means that the actual process that initially created those non-gyrotropic distributions has either stopped or has at least become significantly weaker.

In Fig.\,\ref{F:vel-dist_outflow_high-Q} we highlight some additional regions of enhanced and unexpectedly high non-gyrotropy index $Q_{\rm{e}}$ values in the reconnection outflow and around the secondary peaks in the reconnection electric field $E_{\rm{rec}}$.

We find electrons that are strongly accelerated along the local magnetic-field direction, visible as multiple distinct stripes of enhanced probability at negative $v_{\rm{x}}$; see the rows ``second reconnection'' and ``second acceleration'' in Fig.\,\ref{F:vel-dist_outflow_high-Q}c and d.
This indicates that such electrons were undergoing multiple acceleration processes.
One possible cause is that these electrons were performing multiple meandering motions within the electron diffusion region, which may indeed explain mostly equidistant stripes that are roughly orthogonal to the background magnetic field.
Because we also find a region with an enhanced outwards acceleration at ${\rm{x}} = \pm 3.2$ and ${\rm{z}} = \pm 0.9\,d_{\rm{i}}$, together with a significant reconnection electric field $E_{\rm{rec}}$, we identify a small secondary reconnection site at the location ${\rm{x}} = \pm 2.6$ and ${\rm{z}} = \pm 0.6\,d_{\rm{i}}$.
This finding is underpinned by the fact that the secondary acceleration region is clearly distinct from the main acceleration region that surrounds the reconnection center; see the contour lines in Fig.\,\ref{F:snapshot_overview}.
In earlier works this feature may not have been observed as clearly because of a higher PIC noise level.

Of particular interest regarding the non-gyrotropy are the samples across the separatrix layer that we show in Fig.\,\ref{F:vel-dist_separatrix-shear}.
We also find double-peak shapes (in the \mbox{$v_{\rm{x}}$ -- $v_{\rm{z}}$} distribution) at and below the separatrix layer, which basically comes from an electron velocity shear caused by nearby upstream and downstream flows.
The orientation angle of these double peaks is well aligned with the local magnetic-field vector; see the green line in the ``downstream shear'' row (Fig.\,\ref{F:vel-dist_separatrix-shear}e).

In the region ``current sheet'' we see again stripes mostly parallel to $v_{\rm{x}}$, which is explicable here by multiple meandering motions within the electron diffusion region.
We do not see a significant bulk motion with a negative $v_{\rm{x}}$ here.

When we follow the separatrix layer along the upstream direction, we see strongly non-gyrotropic distributions in Fig.\,\ref{F:vel-dist_upstream} that might misleadingly be interpreted as being close to (or within) the electron diffusion region; see the ``separatrix upstream'' row (Fig.\,\ref{F:vel-dist_upstream}a) with a non-gyrotropy index above $\sqrt{Q_{\rm{e}}} \ge 0.3$.

It is important to note that the $Q_{\rm{e}}$ parameter is indicative of crossing the electron diffusion region only with some additional criterion.
For example, we find that one should also see a distribution with a double peak oriented along the ${\rm{z}}$ direction (or the background magnetic-field gradient) in the \mbox{$v_{\rm{y}}$ -- $v_{\rm{z}}$} components of the electron distribution function (reflecting a meandering motion) in order to identify the electron diffusion region; see the ``reconnection'' regions in Fig.\,\ref{F:vel-dist_reconnection}c, f, i, l, and o.

\section{Discussion and Outlook}

\subsection{Discussion regarding MMS observations}

A complete set of all electron velocity distribution functions within the antiparallel reconnection site (for the original GEM case) discussed in this work is available as a movie online.\footnote{\url{https://doi.org/10.5446/31796}}
One should still note that MMS observations are typically time integrations that represent trajectories through the simulation domain.
Hence, one probably needs to sum up multiple electron VDFs in order to match observations of antiparallel field reconnection.
In follow-up publications we plan to expand this catalog with guide fields, different plasma densities and temperatures, and more realistic mass ratios.

With respect to the recently observed and discussed ``crescent''-shaped electron VDFs \citep[see][]{Hesse+al:2014,Burch+al:2016b} it is worth noticing that we find no such distribution. 
This is expected, because the antiparallel field configuration of this catalog does not fit the dayside magnetosphere.

In this work we find that non-gyrotropic velocity distribution functions for the electrons can occur not only in the electron diffusion region, but also in extended regions, in particular at the electron velocity shear layer close to the separatrix.
Recent MMS observations within the dayside of Earth's magnetosphere revealed non-gyrotropic distributions that are associated with asymmetric reconnection \citep{Burch+al:2016b}.

The MMS mission is going to detect non-gyrotropic distributions also at the nightside of the magnetosphere and one may be misled to a wrong interpretation of the reconnection process because the association between the non-gyrotropic distributions and the spatial regions at or around the reconnection center is difficult.
For an unambiguous identification of the electron diffusion region, we suggest looking for a double-peak electron velocity distribution in the ${\rm{y}}$ -- ${\rm{z}}$ plane along the magnetic-field gradient.
This distribution should also be symmetric with respect to $v_{\rm{z}} = 0$ (where ${\rm{z}}$ is along the background magnetic-field gradient), together with a non-gyrotropy index $Q_{\rm{e}}$ of $0.3$ or higher ($\sqrt{Q_{\rm{e}}} \ge 0.55$).

\subsection{Outlook for solar physics}

Non-Maxwellian electron distributions have been predicted theoretically \citep{Roussel-Dupre:1980} and recently observed \citep{Lee+al:2017} in the solar atmosphere.
Unstable solar magnetic-field configurations (e.g., triggering flares or coronal mass ejections) imply that magnetic reconnection takes place and hence currents exist that may be dissipated to heat the corona \citep{Bourdin+al:2013_overview,Bourdin+al:2014_coronal-loops,Bourdin+al:2015_energy-input}. Magnetic-field parallel electric fields explain the localized acceleration of individual particles \citep{Threlfall+Bourdin:2016_particle-acceleration}.
From this work we see that a ``reconnection-induced current'' is often not a Maxwellian distribution of electrons that is shifted towards the direction of their center-of-mass motion.
Instead, such currents have non-gyrotropic electron velocity distributions caused by the magnetic reconnection processes.
These distributions may feature additional instabilities, within current sheets and when propagating into background plasma \citep{Maneva+al:2016}, which would allow for a better understanding of (or new) onset mechanisms of solar eruptive events.

\subsection{Outlook for future simulations}

This particular work was intentionally performed with a rather simplified PIC setup.
It is obvious that future simulations could be performed with a more realistic mass ratio (at least 10 times larger) and a larger box size allowing for a reconnection that may evolve for longer without influence from any boundary conditions.
Both approaches will result in a significant increase of computational demands.

The GEM parameter settings can be improved with respect to better applicability to the Earth's magnetotail by changing the density, and hence the plasma beta, to more realistic values. Also the influence of weaker and stronger guide fields should be investigated further.

For a better understanding of the plasma-kinetic processes involved, it is a good idea to repeat this experiment while tracking specific particles that resemble certain populations of interest and to inspect their individual trajectories in order to gain insights into the physical processes involved.

Recent 2-D and 3-D kinetic simulations demonstrate that nonsteady turbulent features arise when considering a more realistic large system size and/or 3-D space \citep{Daughton+al:2006,Daughton+al:2011,Fujimoto:2011,Lapenta+al:2015}.
In this study, we did not treat such nonsteady features.
A necessary future research topic would be to provide a catalog including nonsteady regions.

We also suggest adding a much smaller perturbation in the initial condition for similar simulations because this helps to trigger the reconnection more precisely in the box center and allows us to evolve the reconnection more self-consistently.

\subsection{Outlook for future theoretical work}

While we show in our catalog that distribution functions gradually change while we follow the bulk plasma through reconnection, we still find quite characteristic distribution functions for specific locations, like the inflow region, the reconnection center, the acceleration region, and the outflow.
A fundamental physics questions is as follows: can we decompose any distribution found in our model as a superposition of individual transformations that are specific to distinct physical process involved in magnetic reconnection?
In a sense, one could answer this by finding a fundamental and complete set of distributions (or transformations) that would allow us to construct any observed velocity distribution, where one could give ``coefficients'' that represent the influence of each distinct physical process that was involved in forming an observed distribution function.
In return, one would gain insights into which kinetic processes the plasma was undergoing in its history before an in situ measurement.

\dataavailability{
The simulation code, including the changes and input parameters used for this work, can be obtained from \mbox{\url{https://github.com/IWF-Graz/iPic3D/}}; check out the release tag ``GEM-2D\_2016-v1'' and use the ``GEM-smallpert.inp'' input file.
The data from this work may be provided on request.
}

\competinginterests{
The author declares that he has no conflict of interest.
}

\begin{acknowledgements}
I thank Takuma~K.~M. Nakamura and Yasuhito Narita for their helpful suggestions and discussions during this study and Lukas~Lechner for his valuable preparatory work.

The topical editor, Minna Palmroth, thanks two anonymous referees for help in evaluating this paper.
\end{acknowledgements}

\bibliographystyle{copernicus}
\bibliography{Literatur}

\end{document}